# Effects of Defects on the Strength of Nanotubes: Experimental-Computational Comparisons

T. Belytschko, S. P. Xiao and R. Ruoff

Department of Mechanical Engineering
Northwestern University, 2145 Sheridan Rd.
Evanston, Illinois 60208

**Abstract**

The failure stresses and strains of nanotubes given by theoretical or numerical predictions are much higher than observed in experiments. We show that defects can explain part of this discrepancy: for an $n$-atom defect with $2 \leq n \leq 8$, the range of failure stresses for a molecular mechanics calculation is found to be $36\,GPa$ to $64\,GPa$. This compares quite well with upper end of the experimental failure stresses, $11\,GPa$ to $63\,GPa$. The computed failure strains are 4% to 8%, whereas the experimental values are 2% to 13%. The underprediction of failure strains can be explained by the slippage that occurred in the experiments. The failure processes of nanotubes are clearly brittle in both the experiments and our calculations.

Several theoretical and several experimental studies have been made of the strength of nanotubes, although values of the strength have so far only been measured in Yu et al[1]. One of the first theoretical studies of the strength of nanotubes predicted a failure stress of $300\,GPa$ and a failure strain of about 30%[2]. However, it has been shown in Belytschko et al[3] that these large computed values of failure strain and failure stress were an anomaly due to the cutoff function in the Brenner potential that was used in these computations; Shenderova et al[4] have also noted this spurious effect of the cutoff function. Belytschko et al[3] calculated the strength for defect-free nanotubes to be $93\,GPa$ with a modified Morse potential. In addition, they showed that the predicted failure is brittle and that the failure strength depends largely on the inflection point in the interatomic potential; it is almost independent of the separation energy. Neither the results of Yakobson[2] nor the results of Belytschko et al[3] are in good agreement with the experiments of Yu et al[1], where most of the failure stresses were much lower.

Figure 1 shows a histogram of the failure stresses measured by Yu et al[1]. As can be seen, the distribution of the measured failure stresses is very large. Large scatter in failure stresses is also a characteristic of brittle macroscale fracture, where the scatter arises because the initiating mechanisms are small cracks and defects that have large variations in size. Such scatter in macroscale structures is commonly fitted by a Weibull distribution, in which the largest probability of failure occurs at a mean value and the probability of failure decreases exponentially above or below the mean value. However, the measured values of the failure stresses of the nanotubes do not appear to follow a Weibull distribution. Instead, the failure stresses exhibit distinct clusters about a series of decreasing values of strength: the maximum reported failure stress is $63\,GPa$, and there appear to be clusters at $40\,GPa$, $28\,GPa$, $20\,GPa$ and $10\,GPa$. This suggests that the failure of these nanotubes is governed by defects with discrete integer atoms and their strength reduces below the maximum of a perfect nanotube. Because of the small size of nanotubes relative to the atomic scale, any defect would have a much larger effect on the strength than in macroscale structures, so a clustering below the defect-free strength would be expected.

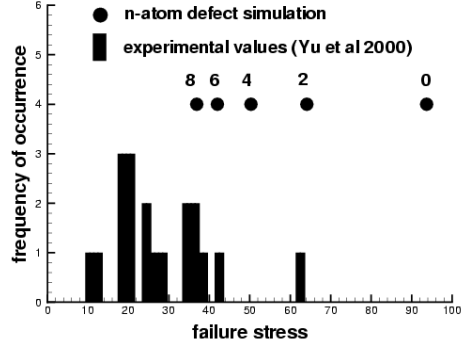

Figure 1. Experimental failure stresses(Yu et al[1]) as compared to computation

In this note, we describe a series of calculations and suggest that the reduction in strength is due to missing pairs of Carbon atoms. The computations were molecular mechanics calculations at $0\,K$. We use a modified Morse potential given in Belytschko et al[3]:

$$E = E_{stretch} + E_{angle} \qquad (1)$$

$$E_{stretch} = D_e \{[1 - e^{-\beta(r-r_0)}]^2 - 1\} \qquad (2)$$

$$E_{angle} = \frac{1}{2} k_\theta (\theta - \theta_0)^2 [1 + k_{sextile}(\theta - \theta_0)^4] \qquad (3)$$

where $E_{stretch}$ is the bond energy due to bond stretch, and $E_{angle}$ is the bond energy due to bond angle. $r$ is the length of the bond, and $\theta$ is the current angle of the adjacent bond. The parameters we used are

$$r_0 = 1.39 \times 10^{-10}\,m,\ D_e = 6.03105 \times 10^{-19}\,N \cdot m,\ \beta = 2.625 \times 10^{10}\,m^{-1}$$

$$\theta_0 = 2.094\,rad,\ k_\theta = 0.9 \times 10^{-18}\,N \cdot m/rad^2,\ k_{sextic} = 0.754\,rad^{-4}$$

This choice of parameters corresponds to a separation (dissociation) energy of 124Kcal/mole. The modified Morse potential is shown in Figure 2 where it is compared with Brenner potential[5].

In the experiments of Yu et al[1], arc-grown multi-walled nanotubes were attached to opposing AFM cantilever tips by a solid carbonaceous material deposited on the tips. In most cases, only the outer nanotube was attached to the cantilever, and only the outer nanotube failed. Therefore, only the outer nanotube was modeled here. The outer nanotubes varied in length from $1.8\,\mu m$ to $10.99\,\mu m$ and their diameters varied from $13\,nm$ to $36\,nm$, so the number of atoms in the outer nanotubes varied from approximately 4 million to 54 million. Models of this size, though feasible, are very awkward so we used [80,0] to [100,0] nanotubes. The nanotubes studied here are significantly smaller than those used in the experiments but we show that the results are almost independent of the size of the model for the defects studied here.

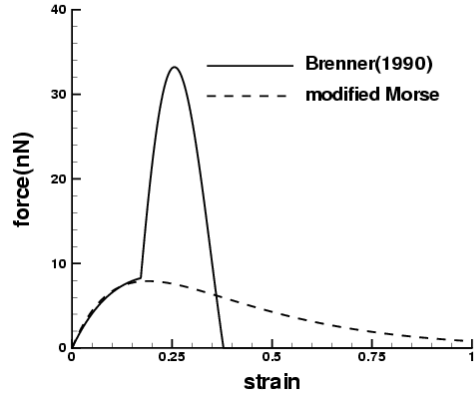

Figure. 2 The force fields for the Brenner and modified Morse potentials

In the simulation, one end of the nanotube was progressively extended until the nanotube failed. In the following we report the stress-strain behavior of the nanotube. The stress $\sigma$ is defined by $\sigma = \dfrac{F}{2\pi r t}$ where $F$ is the axial force, $r$ is the radius and $t$ is a standard thickness used for nanotubes, $t = 3.4 \times 10^{-10} m$. This thickness is the distance between nanotube shells and is a widely used artifact to account for the fact that nanotubes are actually sheets of atoms with no well-defined thickness, see Marino and Belytschko[6] for a theory for the continuum bending stiffness. The strain is defined by $\varepsilon = \dfrac{L - L_0}{L_0}$, where $L_0$ and $L$ are the initial and current length of the nanotube, respectively.

In the simulations, an $n$-atom defect is modeled by removing n adjacent atoms along the circumference of the nanotube. We focus on even values of $n$ in this study because $C_2$ is a more stable form than $C_1$ so that defects involving pairs of atoms is probably more likely. The fracture of nanotube with n-atom defects are studied by using [80,0] zigzag nanotube which consists of 9760 atoms. The dimensions are: radius: $3.1\,nm$ and length: $12.6\,nm$. The failure stresses and strains are listed in the Table I for different defects. These results are also shown in Figure 1 to compare with those of the experiments.

Table I. Failure stresses and failure strains of the [80,0] nanotube with n-atom defects

| defect | failure stress | failure strain |
|---|---|---|
| None | $93.5\,GPa$ | 15.7% |
| 2-atom defect | $64.1\,GPa$ | 8.00% |
| 4-atom defect | $50.3\,GPa$ | 6.00% |
| 6-atom defect | $42.1\,GPa$ | 4.95% |
| 8-atom defect | $36.9\,GPa$ | 4.35% |

Table II Failure stress and failure strain for different nanotubes with 4-atom defect

| nanotube | length | failure stress | failure strain |
|---|---|---|---|
| zigzag [80,0] | 12.6 nm | 50.3 GPa | 6.00% |
| | 16.7 nm | 50.0 GPa | 5.95% |
| zigzag [100,0] | 16.7 nm | 50.5 GPa | 6.00% |
| | 20.9 nm | 50.5 GPa | 6.00% |
| armchair [40,40] | 12.0 nm | 54.3 GPa | 6.40% |
| | 15.5 nm | 53.7 GPa | 6.30% |

To show that the effect of $n$-atom defects on the failure stress and failure strain is independent of the size of nanotubes, [80,0] and [100,0] zigzag nanotubes of various lengths were studied with a 4-atom defect. The radii of the nanotubes were 3.1 nm and 3.83 nm respectively. Table II shows the failure stresses and strains for these nanotubes. As can be seen, the effect of the radius and length on the failure stresses and strains is quite small for a 4-atom defect. Armchair nanotubes are studied here also. As mentioned in Belytschko et al[3], the failure stresses and failure strains of armchair nanotubes are slightly higher than those of zigzag nanotubes.

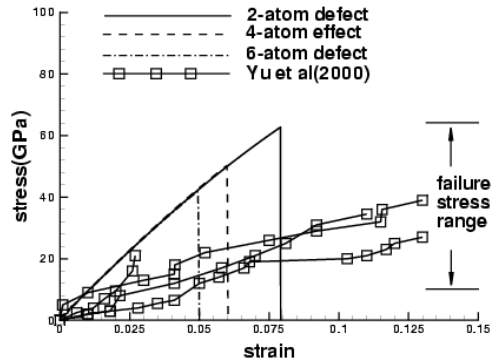

Figure. 3 Computed stress-strain curves compared to the experimental results

Figure 3 shows the stress-strain curve for the [80,0] nanotubes with 2-atom, 4-atom and 6-atom defects as compared to the experimental results. It can be seen that all of the failure stresses for defective nanotubes fall within the experimentally observed range. The failure processes are clearly brittle, with a sudden complete drop in the force carried by the tube. In the molecular mechanics simulations, in the failure process all bonds along the circumference next to the defect broke simultaneously at the critical displacement. On the other hand, in molecular dynamics simulations, the defect initiated a crack which propagated around the circumference of the tube. The axial force dropped to zero in about 0.78 picoseconds.

As can be seen from Figure 3, the failure strains predicted in these studies are significantly lower than those observed experimentally in Yu et al[1]. However, as reported in Belytschko et al[3], re-examination of the experiments led to the conclusion that some slippage may have occurred in the experiments at the AFM tips. Therefore underprediction of the failure strains is to be expected.

The failure stresses are also shown in Figure 1. For the modified Morse potential function, the highest observed failure strength agrees reasonably well with the 2-atom defect and the agreement with the clusters of failure is quite good. This good agreement suggests that our hypothesis about the role of defects in the strength of nanotubes has some merit.

Defects in carbon nanotubes can arise from various causes. Chemical defects consist of atoms/groups covalently attached to the carbon lattice of the tubes like oxidized carbon sites or chemical vapor deposition[7][8]. Topological defects correspond to the presence of rings other than hexagons, mainly studied as pentagon/heptagon pairs[9]. Incomplete bonding defects like vacancies may have been caused through impact by high energy electrons in the TEM environment, see Smith and Luzzi[10] and Banhart[11] or may be defects in the original outer nanotube shell. The thermal conductivity of carbon nanotubes that is dependent of the vacancies has been studied[12]. For a non-stressed single-walled nanotube of diameter $\sim 1.4\,nm$, the atom knockout energy by electron impact has been estimated and also experimentally verified to be of order $85\,keV$. This is significantly below the maximum energy of the electrons in a scanning electron microscope, as used by Yu et al[1]. This suggests the strong possibility that the defects are missing atoms in the outermost shell of these MWCNTs prior to the tensile loading experiments, that is, from the synthesis.

It should be noted that a single $n$-atom defect in the entire outer nanotube suffices to bring about the reduction in strength indicated in Table I. Since the outer nanotubes consists of 4 million to 54 million atoms, the occurrence of a few such defects within a nanotube are quite likely even if the frequency is as low as 1 per 1000 atoms. Furthermore, according to our model, the number of defects will have little effect on strength as long as they are far enough apart so as not to interact. In [80,0] nanotube, there are almost no synergistic or accumulative effects of 4-atom defects if the separation exceeds $8.3\,nm$.

ACKNOWLEDGEMENTS

The support of the Army Research Office and the National Science Foundation are gratefully acknowledged by the first two authors. R. Ruoff gratefully acknowledges the support of the NASA Langley Research Center for Computational Material: Nanotechnology Modeling and Simulation Program.